# SECURITY VISUALIZATION FOR PEER-TO-PEER RESOURCE SHARING APPLICATIONS

DANG TRAN TRI, TRAN KHANH DANG

Faculty of Computer Science & Engineering
HCM University of Technology
National University of Ho Chi Minh City
Ho Chi Minh City, Vietnam
E-mail: khanh@cse.hcmut.edu.vn

*Abstract*— Security of an information system is only as strong as its weakest element. Popular elements of such system include hardware, software, network and people. Current approaches to computer security problems usually exclude people in their studies even though it is an integral part of these systems. To fill that gap, this paper discusses crucial people-related problems in computer security and proposes a method of improving security in such systems by integrating people tightly into the whole system. The integration is implemented via visualization to provide visual feedbacks and capture people's awareness of their actions and consequent results. By doing it, we can improve system usability, shorten user's learning curve, and hence enable user uses computer systems more securely.

**Keywords:** *security visualization; data visualization; peer-to-peer resource sharing; computer security; network security.*

## I. INTRODUCTION

People are increasingly using computer systems for their everyday jobs, from writing documents to chatting with friends, from reading news to electronics shopping. This is largely because of the convenience, powerfulness, cheapness and easily access of these systems today compare to the past. This leads people to more and more relying on these systems and people lives are affected considerably when these systems are malfunctioning or have vulnerabilities. Because computers are usually networked with millions of other computers on the Internet, and these computers are inter-dependent with each other, people are not only depended on machines they own and use, but also on remote servers and communication channels. One of the concerns people have when using computers and Internet for their works is whether their security and privacy are protected. This real concern is anticipated by computer scientists and is the reason behind many advances in information security technology like the use of strong cryptographic algorithms (for example [1], [2] and [3]) to protect sensitive data exchanged via unsafe public networks, or the use of Intrusion Detection Systems (for example, [4]) to detect and notify possible malicious incoming packets. These solutions are proved by mathematical and technological foundations to be effective to protect people resources when residing on their local computers as well as when traveling on intermediate untrusted networks. However, this does not mean security and privacy are protected for people using computer systems as they expected.

As pointed out in [5], people experiences about security and privacy is not always matched with functionalities provided by security products from vendors. For example, people think firewalls not only protect them from unwelcome hackers but also keep unsolicited emails out of their inbox [5]. To actually ensure security for people in their day-to-day activities, these mathematical and technological based solutions are not enough. In addition to these solutions, there must be an emphasis put on the study about the people directly using the system, their computer usage skill level, their knowledge, and their interests. One factor has large impact on people in using a computer system is the user interface. Not all interfaces are intuitive enough for people using them to not have any problems in understanding their action and consequent results. In other term, not all interfaces are usable. To be consider high usability, graphical capability only is not enough, the interface also need to be easy for users to follow and understand; and provide information users need quickly and correctly. Some popular computer system interfaces will be analyzed in term of usability to see why they are not usable from the security perspective. That means they are not easy to use to accomplish what the users want and still guarantee security of the system.

Furthermore, because measuring system usability is about how ease people use them to accomplish their tasks, common people-related security problems are also discussed. This is needed because people is as (if not more) important as technology to ensure system security. For this reason, there is user-centered approach among approaches to design secure software [6]. After discussing about these problems, we then introduce one method to improve secure usability, the visualization technique. This technique has gained much attentions from the researchers and professionals community recently and there are published scientific papers as well as industrial products about enhancing computer system security by visualization ([7] [8] [9] [10]). The focus of this technique is on people who use the system; and it attempt to design an interface that is usable from a security context. To demonstrate the principles presented, we developed a peer-to-peer file sharing application for use in local network environment and we call it LAN P2P File Sharing. This application also helps us in experimenting with various settings to see how visualization affects people and the consequent result on security.

This paper does not attempt to cover all aspects of visualization on security and usability, rather it is meant as an entry point for further studies about security visualization, secure usability and user-center design for security. The rest of this paper is organized as follows: Section II briefly introduces fundamental security problems and existing solutions related to people. Section III presents solutions employing visualization to enhance the system security. Section IV introduces our improvements, based on security visualization techniques, for peer-to-peer resource sharing applications. Section V gives concluding remarks and presents future work.

## II. PROBLEMS WITH THEORETICAL-BASED SECURITY SOLUTIONS

Theoretical-based security solutions work on the assumption that their users will act correctly according to their need. For example, public-key cryptosystems work on the assumption that their users can successfully keep their private keys secret, or a file sharing systems work on the assumption that their users will select "Read" when they want others to read their files. However, as we will discuss, these assumptions are not always true.

The truth is people may or may not understand the security mechanism behind these solutions, and they also may or may not have motivation enough to learn more about that. Usually, people buy a security product simply because they want to be protected in general [5]. They do not know which resources can be protected by a particular product and which is not. Instead, people seek for an all-in-one solution that can make them safe as a whole [5]. This is clearly impossible. People should be made aware that their actions contribute much to the security of a system, maybe more than that of security solutions. For example, an authentication system can not provide any useful access control functionalities if the users of it disclose their passwords to others; or a public-key cryptographic system can't protect its user's data once their private keys are shared to the world, instead of public keys, by mistake! Positive people contribution to security can only be improved by educating people about product features and making user interfaces easy to use and difficult to make mistakes. However, not all people want to invest their time and efforts in learning more about security, so it's a system designers' responsibility to design effective user interfaces for security purposes. We will now consider the relationship between security and usability of computer systems, and then discuss some problems of security that is people-related.

A. **Security and usability**

Usability of a program is measured as how easy it is for people using it to accomplish their particular tasks. However, it should be noted that usability may have different meanings in different contexts. For example, one may view usability of a program as how productivity people can use it for their works, and others may regard its usability as how easy people can learn to use it fluently, etc. From a security perspective, we use the definition of usability for security software from [11] as follows:

*Security software is usable if the people who are expected to use it:*

1. *are reliably made aware of the security tasks they need to perform;*
2. *are able to figure out how to successfully perform those tasks;*
3. *don't make dangerous errors; and*
4. *are sufficiently comfortable with the interface to continue using it.*

In this paper, we will generalize the above definition about usability for security software to any computer systems that have to deal with security at some degree (which we believe almost, if not all, systems have). Note the four points of the definition above are valid not only to security software but also to any systems that have a security subsystem. So, we make these two small modifications to the definition: instead of "security software", we replace with "computer systems", and instead of "usable" we replace it with "secure usable". With these modifications, we mean any computer systems can be considered usable from a security point of view if they have four defined attributes above. For example, with a browser, it is not considered securely usable if the people using it can not easily differentiate between cases where their submitted data is protected and cases where it is not; or for a file sharing program, if its interface allows users to easily select "Delete" shared permission by mistakes although they just want others to read their files, it is not considered securely usable either.

An unusable system not only makes it's hard for authorized people using it, but also leads to security threats as these people may find ways to bypass the restrictions to accomplish their tasks easier, ways that system designers may not expected before when designing the system. Because this is not clear at first thought, we will illustrate with an example: On some computer systems, there is a password restriction (or password policy) placed on users' accounts to protect their identities. These systems require users creating their passwords with restrictions like: the password length must be longer than a minimum length; combination of upper case, lower case, number, and special characters is a must; and password is only valid for a period of time, after which users need to change them to new passwords, etc. Although that makes users' passwords more secure and hard to guess, it also creates much difficulties for casual users in choosing and remembering their passwords. Due to this obstacle, some users may write down their passwords somewhere, which in turn makes the system not secure anymore.

But a usable system does not guarantee the security of it. There is a trade-off between security and usability that is mentioned by experts in the field. For example, on November 2000, in his *alertbox*, Jakob Nielsen, who is called "The king of usability" by Internet Magazine, said that [12]:

– *Usability advocates favor making it easy to use a system, ideally requiring no special access procedures at all, whereas*
– *Security people favor making it hard to access a system, at least for unauthorized users.*

For some applications, usability is placed at a higher priority than security, and for others it is reversed. Some applications even let people adjust this according to their needs. For example, popular browsers like Internet Explorer or Mozilla Firefox have the feature of "Remembering password" on authentication forms for convenient using of people. However, people will decide whether they want more security (by not letting the browsers remember their passwords) or more usability. But not all applications provide this kind of flexibility, instead they provide a default behavior that the system designers think the users will response with high possibility (but of course, it may or may not match with users' actual thought). One particular example is the Windows Vista alert message box when users delete a file:

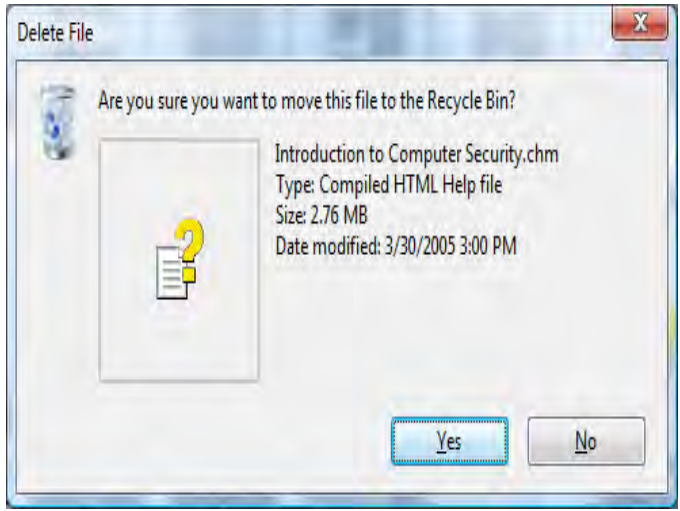

**Figure 1: Windows Vista delete file message dialog**

The default button in this message box is "Yes" button. Although this interface design is not very secure because users can easily delete a file by mistake, but it is more usable from a productivity perspective.

Another difficulty in dealing with usability for security as described in [11] is that designing an effective interface for security purpose is not the same as designing interfaces for other types of product. That is because effective security requires a different usability standard. The paper [11] described an experiment with the interface of one popular security software: PGP 5.0. This software interface is considered as good usable from a general consumer software standard. However, experiment results showed that it is not usable from the security perspective because its users could not easily accomplish their security tasks. This leads to the need to have researches about system usability for security. Now, there is a new research area named HCI-SEC (Human Computer Interaction and Security) study about the relation between usability and security and how to improve both of them at the same time.

**B. People-related security problems**

Security solutions are only effective when the people using it correctly. Usually, for people, protecting their security and privacy is not as critical as getting their jobs done. They just want to accomplish their tasks with least possible effort even though that may mean bypassing the security system. System designers should take into account that issue and do not assume users will try their best to make sure security and privacy are protected or users will spend much time reading manual to know how to use the system securely. As [13] demonstrate, experiments showed that security browser tool bars fail to protect their users from phishing attack and one of the reasons is because users usually don't care about warning messages these tool bars provide. They care more about content and look-and-feel of visited sites and hence are deceived easily by average attackers (just need to save original website content and upload to attacker's site).

In case the person guarding a system it not the person suffers when that system fail, the situation is worse [14]. From the economic point, it is not that these guarding people don't want to secure their systems, just because the design and deployment of such systems do not match with people's incentive. For example, the owner of a zombie computer may not know or care that the computer will be used to attack against other machines because it does not make any harm to this person. But if the computer is infected by a virus with destroying behaviors, this person will be much critical worry.

There is a research area call Incentive Centered Design (ICD) with the definition: ICD has the aim of designing systems that respect motivated behaviors, by providing incentives to induce human choices that improve the effectiveness of the system [15]. When this approach is used broadly in designing security systems, the security threats by misaligned incentives will be reduced significantly.

Not only people using a system are important for the security of that system, attacker's behaviors also play an important on that. Again, the relation between attackers and defenders can be explained by economics theory. But in summary, according to [16], attackers of computer systems usually are more motivated in their actions than defenders and also have some advantages over defenders. This can further make the job of system designers more difficult in designing a secure system not only from the technical side, but also from the human psychology and economy sides.

## III. USING VISUALIZATION TO INCREASE SECURITY

Although computers excel at processing standard data, it is beaten by human in recognizing and analyzing novel patterns. This is especially true in computer security field, in which analytic capabilities and creativity of humans place an important role in keeping the system safe. But to take advantages of human capabilities in recognizing and analyzing these patterns, it is necessary that these data is presented in a format that's easy for human to grab quickly as well as spot the abnormal easily. One way to do this is using visualization.

Visualization is a technique used to present data in pictorial forms. This is particularly useful as people often cite "A picture is worth a thousand words" (Raffael Marty in his book "Applied Security Visualization" also said "A picture is worth a thousand log records" [17]). A picture may provide needed information quickly compare to description in text form. For example, a map definitely helps travelers find their way to their interesting destinations easier and quicker than the description text on how to get to these places. Not only helping people in understanding data, visualization may also help people in predicting future situations based on the patterns presented. One example is the stock price charts, which knowledgeable people can use to predict if the prices will rise, fall or stand. This can also be seen in business applications, which usually provide the simulation and visualization capabilities to decision makers so they can easily experiment with different inputs data.

At first thought, visualization may be considered similar to computer graphics in which both are used to present data in image form. However, the difference lies in the different purposes of them.

While computer graphics is used to present an experience as real as possible to user, visualization is not just used to present data. As it was conceived in 1980s, visualization is used as an interactive process to understand what produce data, not a method to present it [18]. Because visualization is mean for people to understand the process which creating it, it is important to place people in the center of designing visualization models task. If a model is not easy for people to understand, it is not considered a good model.

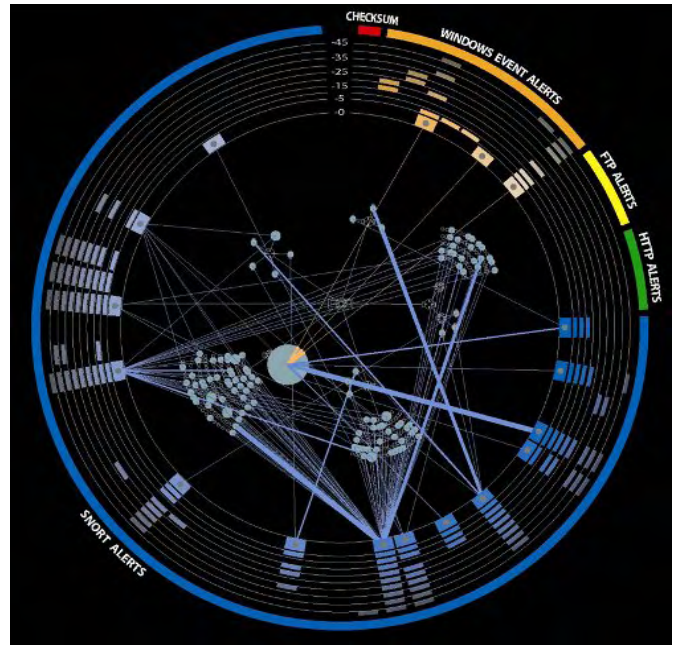

**Figure 2: VisAlert interface (image from [19])**

In computer security, visualization can help people in using the system more effectively by helping people in understanding the underlying data without the need of investing much effort. Especially in cases where the source data is very large and complex, for example, alert logs created by an Intrusion Detection System (IDS). In these cases, visualization can be used to selectively provide only essential information to people of the system in forms that is most easily captured by human cognitive system. For example, [19] described a method for visualizing alert messages. The tool is used to demonstrate the principles in this paper is called VisAlert, which use visualization to present alerts on separated logs created by an IDS. VisAlert display 3 essential information that each alert has to contains: what, when and where about an alert to which administrators can quickly view to know about alert types, time happen, and at which machine it occurs respectively. Administrators can also guess quickly with greater confident about whether some alerts are false or true from the visualization model presented once they use it long enough and hence getting some real experience predicting alerts' validity.

The work of [9] described another visualization technique called Binary Rainfall which can be used to analyze binary data to detect malicious binary objects. It can be used to compare 600-1000+ objects at one time, thereby increasing productivity of the administrators using it. Another popular visualization technique is *treemap* [20], which is an effective visualization technique to display large hierarchical data and there are many applications of it in different fields. A particular application of the treemap in security is discussed in [21], that can help people to have an understanding of global permission configuration of the large file systems quickly.

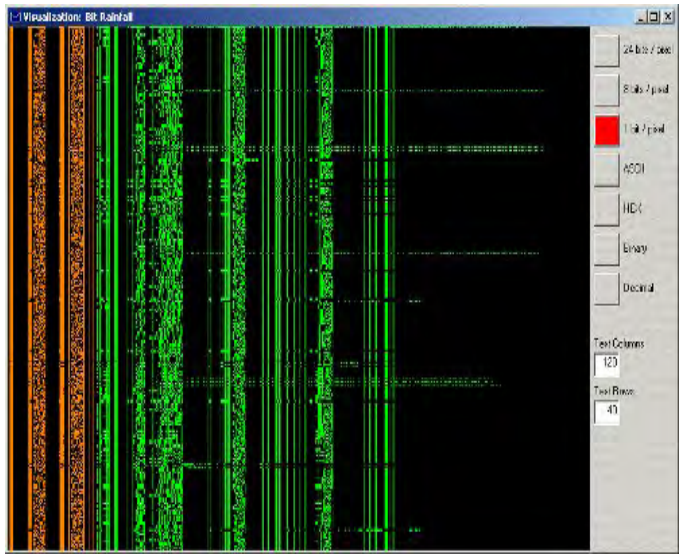

**Figure 3: Binary Rainfall Visualization of Defcon 11 "Capture the Flag" Network Traffic (image from [9])**

Not only helps people in understanding underlying data, visualization is also used to help increase usability of a system and at the same time increase its security as well. To accomplish that, visualization techniques are implemented to provide instant feedback about people's actions, thus make people more aware about the security implication of their actions.

For example, Impromptu client interface [22] uses visualization to display system states, and integrating configuration and action into one screen. By using different colors and positions to present different users and shared files, people can easily know which file is owned by who and monitor each user activities. And because configuration and action are now integrated, people can execute their actions and view global configuration states result at the same time. That will make people more aware about their security-related actions.

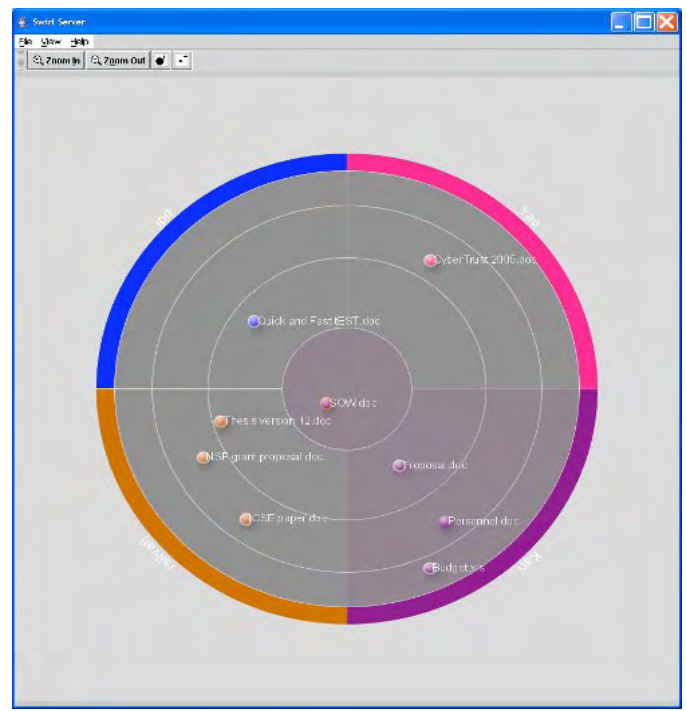

**Figure 4: Impromptu client interface (image from [22])**

Other than providing useful feedback to improve secure usability, another technique to improving usability is reducing the complexity involved for decision makers. For example, Windows XP provides an interface to set permissions on file and folders to users and groups. However, it does not contain group information for a specific user or user information for a specific group. This issue plus rules on permissions priority and combination make it very hard for administrators to know what effective permission a user has on a folder or file, not to mention about permission created by hierarchy and inheritance relationship between folders/files in the file system which further duplicate the problem. The work in [23] discuss one possible way to solve that problem by providing an alternative interface for setting files' permissions in Windows XP environment with much more usability. This program provides all the needed information about users, groups, permissions, and final effective permission on one screen, so administrators do not need to access many places to do the checking on individual settings and calculating effective permissions. With this enhancement in usability, this program improves correctness of configurations on files and folders permissions created by administrators and overall increase security of the system.

handling with data ranging from overview to detail. In [7], [8], the authors use visualization to display network activities. In its initial setting, it works in "galaxy view", which provides an overview of whole class B network activities. However, it's an interactive application and people can choose the scale at which to display based on their need. For example, it provides a "small multiple view" to display a small selected network portion or "machine view" to view network data goes to and from a particular machine.

## IV. LAN P2P FILE SHARING

This program is developed to experiment with visualization to improve security of a popular application: sharing files on local network with each other. Each shared file will have different share modes: read – others can copy this file to their own machine, write – others can "read" and copy their local machine file to replace this file content, and full – other can "write" and delete this file from the owner machine as well. In addition to share file, each user may have the need to see what happen to their shared files. We use the same interface for both sharing file and providing events information happen to shared files.

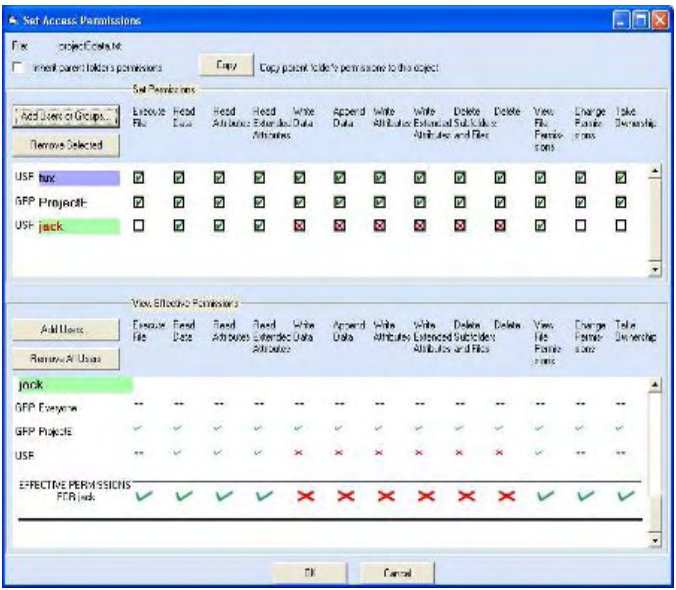

Figure 5: Salmon interface (image from [23])

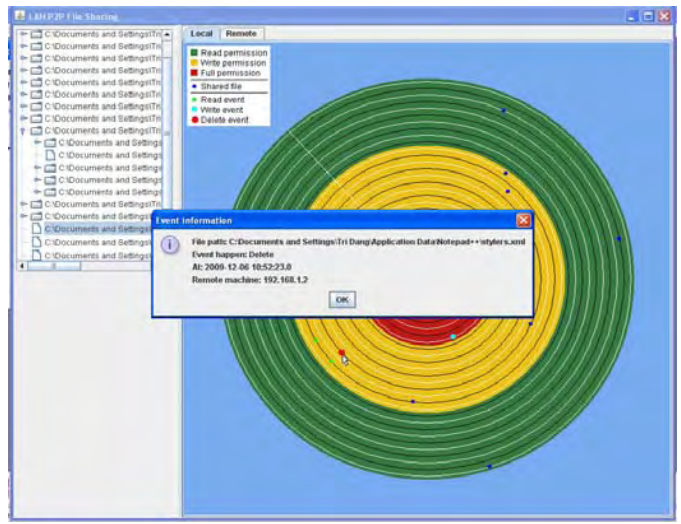

Figure 6: LAN P2P File Sharing let users see events happened on their systems and share files at the same time

Visualization can include the abilities for people to zoom to a particular level of interest, or filter to see only needed data only, to name a few. These features can make visualization a powerful tool in

This is different from Windows XP, which in order to know happened events, users need to open another program called "Event Viewer" to view event logs. This design will make users see important events to their file so they can make appropriate actions. Removing the need to go to another screen just to see happened events, the application provides useful information even for unmotivated users. Because users' primary objective here is sharing file, so if the event information is on another screen that may make users do not bother to have a look at them at all.

Secure usability of the program is implemented via the following techniques: Provide instant feedback about users' actions. Whenever users move their local files to sharing locations, they will be provided instantly with information about what can happen with their files as Figure 7 depicted.

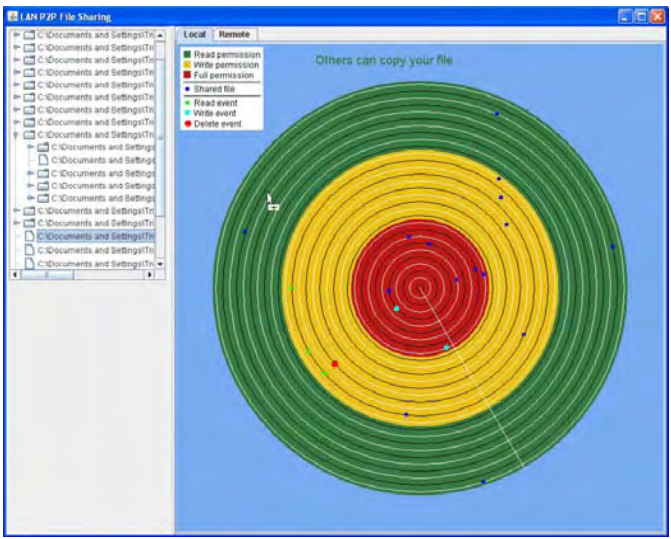

**Figure 7: A human-readable description appears at the top-center of main interface to help new users familiar with the program quickly**

## V. CONCLUSION AND FUTURE WORK

Improving security of a computer system requires improving all of its elements because security of an information system is only as strong as its weakest element. It is well-grounded that people are the weakest link in security [24]. Although security technologies advance quickly, it is not easy to educate people how to use computer securely. That means to improve security we must improve user interaction with system by providing effective interface for them. As this paper demonstrated, visualization techniques can be used to improve user involvement in security process by providing visually useful feedbacks and integrated screens for a variety of related actions.

Even then, visualization is not a total solution to people-related security issues, and it can also give the false information to users once attacker know how to attack visualization system by overloading information and deceiving human cognitive [25]. To be highly effective, visualization systems need to work with knowledgeable users. Of course not all people have motivation to learn about security, but for motivated people, effective teaching methods should be used when possible to get the most from these learners' time. Some methods about security teaching are discussed in [26], and this interestingly challenging research topic is also of great interest for our future work. Besides, in the future, we are also going to carry out intensive experiments on real world applications in order to establish the practical value of our proposed method.

### ACKNOWLEDGMENT

We are thankful to ASIS Lab members (Advances in Security & Information Systems Lab at the Faculty of CSE, HCMUT) for their enthusiastic help and support during writing this article.